\documentclass[12pt,preprint]{revtex4}
\usepackage{amsmath}
\usepackage{amssymb}
\usepackage[dvips]{graphicx}
\begin{document}
\title{Parameter Estimation by Density Functional Theory 
for a Lattice-gas Model of Br and Cl Chemisorption on Ag(100)}
\author{Tjipto Juwono, Ibrahim Abou Hamad, Per Arne Rikvold}
\affiliation{\it Department of Physics, Florida State University,\\
Tallahassee, Fl 32306, USA}
\author{Sanwu Wang}
\affiliation{\it Department of Physics \& Engineering Physics,\\
University of Tulsa,\\
Tulsa, OK 74104, USA}
\begin{abstract}
\noindent 
We study  Bromine and Chlorine chemisorption on a Ag(100) surface, using a lattice-gas
model and the quantum-mechanical Density Functional Theory (DFT) method. In this model the Br and
Cl ions adsorb at the fourfold hollow sites of the Ag(100) surface, which
can be represented by a square lattice of adsorption sites. Five different coverages  were used for each
kind of adsorbate. For each adsorbate and coverage,  we
obtained the minimum-energy configuration, its energy, and its charge distribution.
From these data we calculated dipole moments, lateral interaction energies,
and binding energies. Our results show that for Br the lateral interactions 
obtained by fitting to the adsorption energies obtained from the DFT
calculation are
consistent with long-range  dipole-dipole lateral interactions obtained using the
dipole moments 
calculated from the 
DFT charge distribution. For Cl we found that,
while the long-range dipole-dipole lateral interactions are important, 
short-range attractive interactions are also present.
Our results are overall consistent with parameter estimates previously 
obtained by fitting room-temperature Monte Carlo simulations to 
electrochemical adsorption isotherms 
[I.\ Abou Hamad et al., J.\ Electroanal. Chem.\ 554 (2003), 211; 
Electrochim.\ Acta 50 (2005), 5518]. 
\end{abstract}
\date{\today}
\maketitle
\section{Introduction} 
\noindent 
The adsorption of halides on noble metals provides important model systems for studying adsorption on metal surfaces, particularly when there are ordered adsorbate structures. For this reason, these adsorption processes have been extensively studied~\cite{ock,ock1}. Adsorption of Bromine and Chlorine on metal has been the 
subject of many studies over the years. The systems we study here are Br and Cl chemisorbed on single-crystal Ag(100).
Experimentally, 
Kleinherbers  $et~al.$~\cite{klein} have found that the adsorption 
of Br, Cl, and I on Ag(100) surfaces in vacuum all resulted in the 
formation of a $c(2 \times 2) $ overlayer with the adsorbates in the 
fourfold hollow sites. 
This implies a very strong, short-range repulsion, which we model as a 
nearest-neighbor exclusion \cite{kop}. 

The bonding of the adsorbates to the substrate and the surface electronic 
structures have been studied by Density Functional Theory (DFT) calculations. 
It is found that the bond between Br or Cl and the substrate is covalent with a polarization due to  electron 
transfer from the substrate to the adsorbate~\cite{sanwu,kramar,rik}.  The polarization results in dipole moments on the surface, which cause long-range dipole-dipole interactions between the adatoms.

Long-range dipole-dipole interactions have previously been incorporated in a 
lattice-gas model employed in room-temperature 
Monte Carlo simulation studies of the adsorbed system~\cite{hamad:2,hamad:1}. 
In these works,  the lateral interactions were extracted by fitting the 
results of the simulations to electrochemical adsorption isotherms.  
In the present study we instead estimate the lateral interactions 
by fitting the lattice-gas model to our DFT results.

We extract the next-nearest-neighbor lateral energy and the binding energy by fitting the lattice-gas model to the adsorption energies obtained from 
the DFT calculation. The same DFT calculation also yields charge 
distributions from which dipole-dipole interactions can be directly 
calculated.  By comparing the two results,  we examine the significance of 
the long-range dipole-dipole interactions within the lattice-gas model.

In this study we present DFT calculations using supercell models for 
the Ag(100) surfaces. The adsorbates in these DFT calculations are assumed 
to occupy a lattice of adsorption sites in accordance with a lattice-gas 
approximation~\cite{mit}. 
The lattice-gas assumption of strongly located adsorbates is consistent with 
previous DFT calculations and dynamic Langevin-equation simulations for 
a continuum model \cite{rik}.

The adsorption energies and charge distributions were calculated by DFT. We assume long-range dipole-dipole 
interactions between the adsorbates,  and we implement these long-range 
interactions in the fitting of the lattice-gas model to the adsorption-energy 
results and the dipole moments obtained  from the DFT calculations.

Estimates of short-range lattice-gas interactions from DFT calculations of  
adsorption energies have also recently been performed for 
homoepitaxy \cite{STAS06,TIWA07,LIU10} and 
heteroepitaxy \cite{LIU10} systems. However, these studies do not consider 
charge transfer and long-range interactions. 

The rest of this paper is organized as follows. Section 2 describes the 
details of the DFT calculations and the methods used to analyze the results, 
Section 3 discusses the calculation of the dipole moment, 
Section 4 discusses the lattice-gas model, 
Section 5 presents the lattice-gas fitting, 
and Section 6 contains a discussion.

\section{Density Functional Theory}
\noindent We applied DFT to obtain
the ground-state energies  and  electron density functions for 
the adsorption of Br or Cl on a slab representing a Ag(100) surface. 
We prepared slabs with seven metal layers. Convergence checks with respect to the number of layers are discussed in Appendix A. The slab was placed inside a supercell with periodic boundary conditions. Two different sizes of supercells were examined. 
A $2 \times 2$ supercell with the size of $2a \times 2a \times 37.53 \rm\AA$ , and a  $3 \times 3$ supercell with the size of $3a \times 3a \times 37.53\rm\AA$. Here, $a=\alpha/\sqrt{2}$ where $\alpha=4.17 {\rm\AA}$ is the lattice 
constant of bulk Ag, which we obtained from DFT calculations by minimization of an Ag fcc structure.
The $2 \times 2$ supercell contained four surface Ag atoms on each side of 
the slab (28 Ag atoms in total), while the $3 \times 3$ supercell contained 
nine surface Ag atoms on each side (63 Ag atoms in total). 

\begin{figure}[t]
\scalebox{0.5}
{\includegraphics{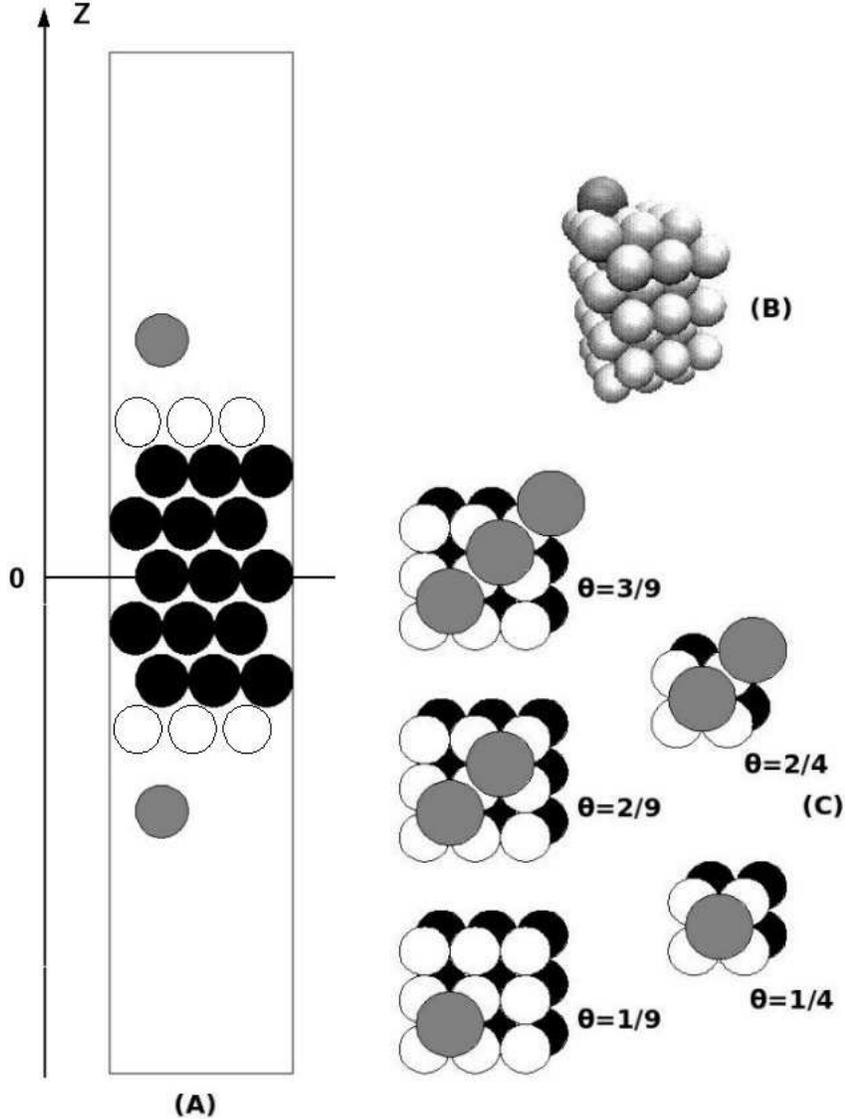}}
\caption{(A) The cross section of the  supercell, (B) a three-dimensional 
representation of the supercell,  and (C) surface distributions  of 
the adsorbates for various coverages. Adsorbate atoms: gray. Surface Ag atoms: white. Bulk Ag atoms: black.}
\label{super}
\end{figure}

The orientation of the surface normal defines the $z$ direction. 
To maximize the symmetry, we distributed the adsorbates on both sides of the 
slab. One, two, and three Bromine or Chlorine atoms were placed on each 
$3 \times 3$ surface to represent coverages $\theta= 1/9$, $2/9$, and $1/3$, 
respectively. Two Bromine or Chlorine atoms were placed on each 
$2 \times 2$ surface to represent $\theta=1/2$  and one to 
represent $\theta = 1/4 $.
 Here the coverage $ \theta $ is defined as
\begin{eqnarray}
\theta=\frac{1}{N_{\rm site}}\sum_i c_i,
\label{cov}
\end{eqnarray}
 where $ c_i = 1 $ when the site is occupied by the adsorbate, and $ c_i=0 $ 
otherwise.  In other words, the coverage is the number of adsorbates 
divided by the total number of all possible adsorption sites, $N_{\rm site}$.  
Figure~\ref{super} shows the cross section of a supercell and surface 
distributions of the adsorbate for various coverages. Due to the 
nearest-neighbor exclusion and the periodic boundary conditions the adsorbates can only be placed in diagonal positions, limiting $\theta$ to less than or equal to $1/2$.

The calculations were performed by the DFT method using the Vienna Ab Initio Simulation Package (VASP)~\cite{kresse:3,kresse:2,kresse:1}. The basis set was plane-wave, with the generalized gradient-corrected exchange-correlation functional~\cite{perdew:2,perdew:1}, Vanderbilt pseudopotentials~\cite{vanderbilt:1,kresse:4}, and a cut-off energy of 400 eV. The $k$-point mesh was generated using the Monkhorst~method~\cite{mon:1} with a $7\times 7 \times 1$ grid for the $2 \times 2$ supercells and a $5 \times 5 \times 1$ grid for the $3 \times 3$ supercells.
To get to the configuration with minimum energy, we used a selective dynamics method, by which the ions 
in the top and bottom layers were allowed to relax in the $z$ direction only,  as opposed to the full dynamics in which the atoms would be  allowed to move in all directions. This is the first step to avoid surface reconstruction, 
which is not expected to occur in this system under electrochemical conditions. The second step is to average the $z$ coordinates of the top and bottom layers.
The DFT results yield total energies and electron densities, $\rho_e(\vec{x})$. 

We next ran static minimization on the resulting averaged minimum-energy structure. Here, `static' means running energy minimization on the electron distribution without changing the positions of the nuclei. From this run, we obtained the total energy of the system, $E_{\rm{syst}}$. We then took the same structure and removed the adsorbate to obtain the clean-slab structure. 
Again, we ran selective dynamics on this slab structure to 
obtain ${E_{\rm{slab}}}$. To get the energy of an isolated halide atom, we also ran static minimization on an isolated halide atom to obtain $ E_{\rm hal} $.  We define the adsorption energy per supercell per site as

\begin{eqnarray}
E_{\rm ads}=\frac{E_{\rm syst}-E_{\rm slab}-2NE_{\rm hal}}{2N_{\rm site}}.
\label{eads1}
\end{eqnarray}

\begin{figure}[t]
\scalebox{0.6}{\includegraphics{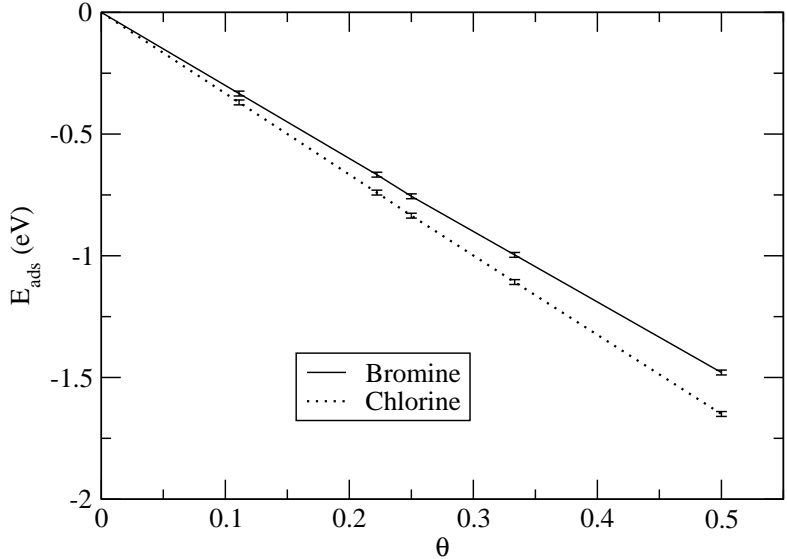}}
\caption{Adsorption energy vs coverage. Despite the appearance on this scale, the lines are, in fact, slightly convex due to the repulsive lateral interactions. The ``error bars'' in this and subsequent figures do not represent statistical errors, but rather estimates of the accuracy of the results, based on the convergence studies discussed in Appendix A.}
\label{adse}
\end{figure}

\noindent Here, $N_{\rm site}$ is the number of sites on one surface of the metal slab, and $N$ is the 
number of halides on each side of the slab. 
In Fig.~\ref{adse} we show $E_{\rm ads}$ as a function of $\theta$ for both 
systems. We emphasize that $E_{\rm ads}$ contains the lateral interaction energy and is different from the single-particle binding energy $E_{\rm b}$. The relation between these two quantities is given explicitly in Eq.~(\ref{eads}) below.

To understand the surface polarization  we need to study the charge-transfer behavior.  We define the negative of the electron densities from the DFT output as the charge density distributions $ \rho(\vec{x}) $,  and we introduce the charge transfer function per adsorbed atom,  which is defined as follows~\cite{mit:brau}

\begin{equation}
\Delta\rho(\vec{x}) = \left[ \rho(\vec{x})_{{\rm halide}-{\rm Ag(100)}} - \sum_{i=1}^{N}\rho(\vec{x})_{\rm halide}-\rho(\vec{x})_{\rm Ag(100)} \right]/N ,
\label{surface_dip0}
\end{equation}
where $\rho(\vec{x})_{\rm halide-Ag(100)}$ is the full charge density of the adlayer system with $N$ adsorbed Br or Cl on each side of the slab, and $\rho(\vec{x})_{\rm halide}$ is the full charge density of the pair of isolated halide atoms at the same positions as  in the halide-Ag bonded system, and $\rho(\vec{x})_{\rm Ag(100)}$ is the charge density of the Ag(100) slab with all atoms at the same positions as in the halide-Ag bonded system~\cite{leung}.
After integrating over $x$ and $y$, this yields the charge transfer function per pair of adsorbed atoms,

\begin{equation}
\Delta\rho(z) = \lbrack \rho(z)_{{\rm halide}-{\rm Ag(100)}} - N\rho(z)_{\rm halide}-\rho(z)_{\rm Ag(100)} \rbrack/N.
\label{surface_dip4}
\end{equation}

\begin{figure}[t]
\scalebox{0.6}{\includegraphics{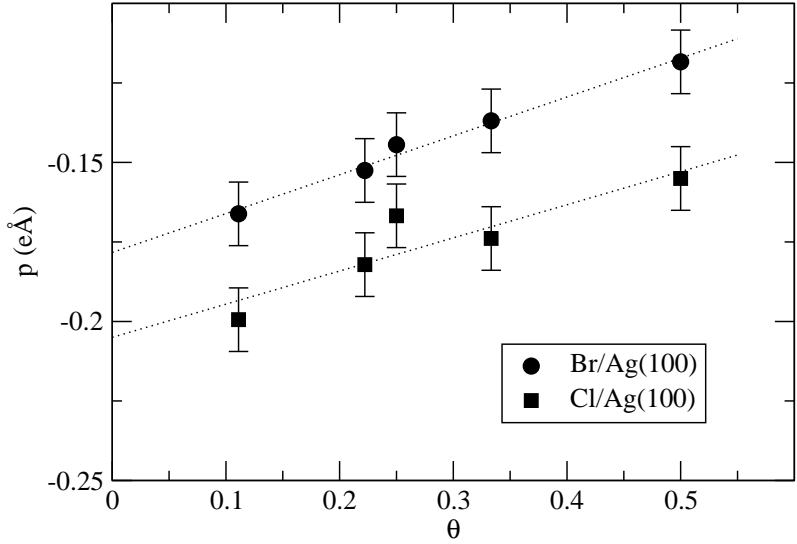}}
\caption{Dipole moment vs coverage.}
\label{dipm}
\end{figure}

From the charge transfer function  integrated over the $x$ and $y$ directions, $\Delta \rho(z)$, we can calculate the surface dipole moment as  
\begin{equation}
p= \frac{1}{2}\int_{-h}^{+h} \left| z \right| \Delta \rho(z)dz.
\label{surface_dip}
\end{equation}
\noindent Here $h=\frac{1}{2}H$ where $H$ is the height of the supercell. 
The zero point of the coordinate is placed at the middle of the supercell. 
Figure~\ref{dipm} shows the results of the dipole moment calculation for 
Bromine and Chlorine. Here we observe that the magnitude of the dipole moment decreases approximately linearly with $\theta$.  The surface dipole moment of 
the energy-minimized clean slab was also calculated and verified to be 
the same as the surface dipole moment of the slab with all atoms at the same positions as in the halide-Ag bonded system, thus justifying our procedure. 

\begin{figure}[t]
\scalebox{0.6}{\includegraphics{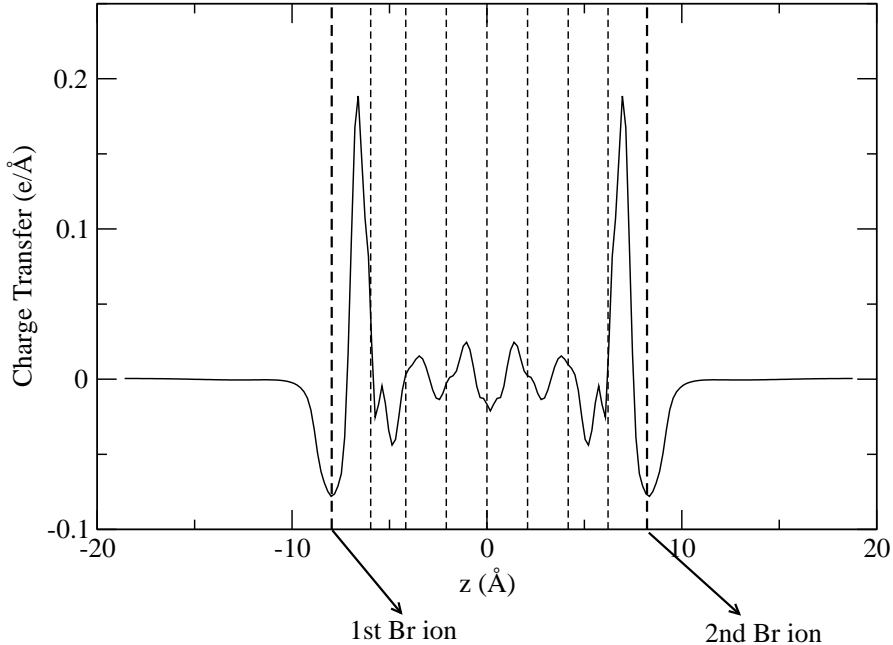}}
\caption{The charge transfer function $\Delta\rho(z)$ for 
Br/Ag(100) with $\theta=1/9$. The vertical short-dashed lines indicate the $z$-positions of the Ag layers, 
and the long-dashed ones indicate those of the adsorbate ions.}
\label{layers}
\end{figure}

Figure~\ref{layers} shows the charge transfer function $\Delta \rho(z)$ for Br/Ag(100) with $\theta=1/9$.  
In this figure positive values indicate electrons being removed, while 
negative values indicate electrons being added.  From the figure we see that charge is mostly transferred  from the surface silver atoms to the adsorbates. Inside the bulk, 
the charge transfer function indicates only minor charge redistribution above and below each of the  silver layers. Since the charge transfer function is calculated by subtracting the charge distributions of the clean slab and isolated adsorbate from that of the adsorbed system, we conclude that this small charge redistribution is caused by the adsorption processes.

\begin{figure}[t]
\scalebox{0.6}{\includegraphics{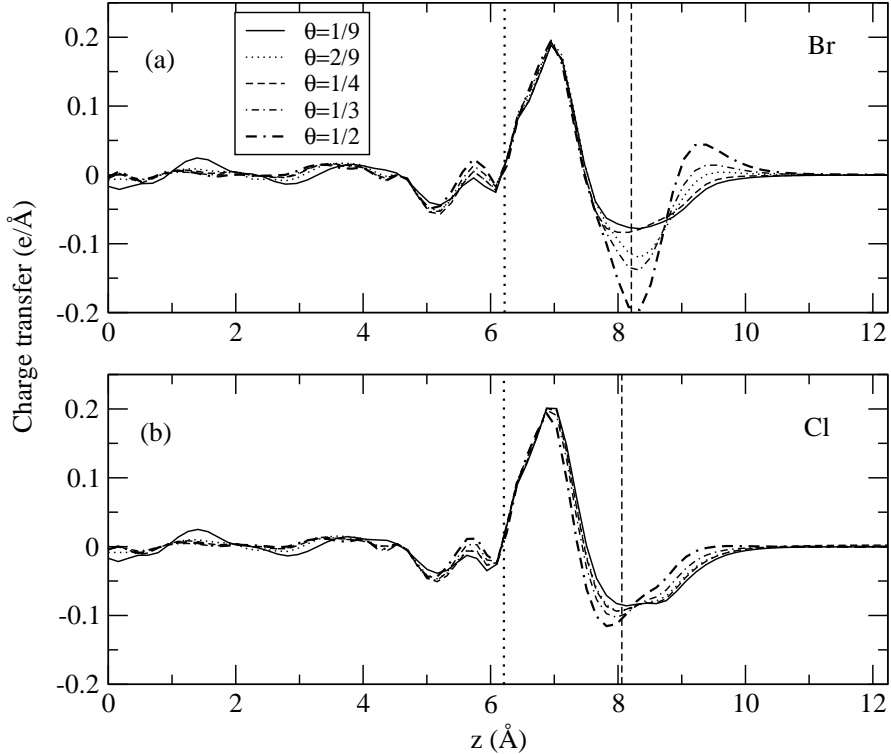}}
\caption{Charge transfer functions $\Delta\rho(z)$ for all coverages. Only half of the supercell is shown, from $z=0 \rm \AA$ to $z=12.235 \rm \AA$. Panel (a) is for Br/Ag(100) and (b) for Cl/Ag(100). The dotted lines correspond to the $z$-position of the topmost layer of metal and the dashed lines correspond to the $z$-position of the adsorbates.}
\label{allcov}
\end{figure}

\begin{figure}[t]
\scalebox{0.6}{\includegraphics{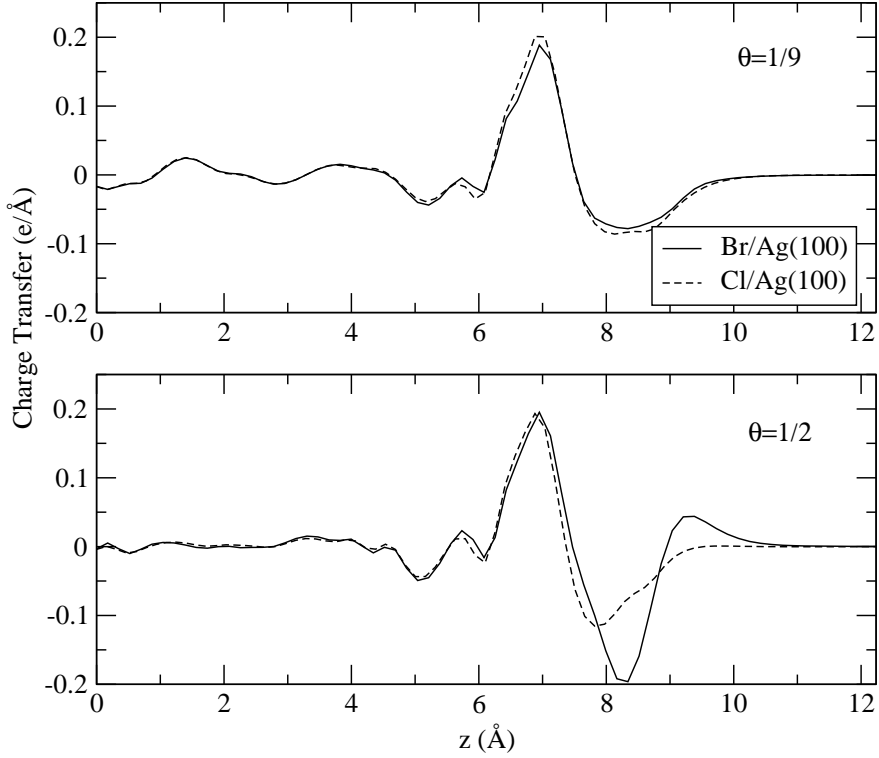}}
\caption{Comparison of charge transfer functions $\Delta\rho(z)$ for for Br and Cl at 
two different coverages.}
\label{brcl}
\end{figure}

Figure~\ref{allcov} shows the charge transfer function per adsorbed atom $\Delta \rho (z)$ for Br/Ag(100) and Cl/Ag(100) for all coverages. Here we only show half of the supercell since the charge transfer function is symmetric in the $z$ direction. Both systems show a similarity in that the magnitude and distribution of the charge transfer from the Ag surface  are independent of the coverage.   
However, Fig.~\ref{allcov} also shows that while the magnitude of the charge transfer from the surface to the adsorbate is independent of the coverage, the resulting charge distribution  around the adsorbate is not. 
Indeed, higher coverage results in a more asymmetrical charge distribution around the adsorbate. This asymmetry is more pronounced in the Br/Ag(100) case, 
suggesting an important difference between Br/Ag(100) and Cl/Ag(100).  Figure~\ref{brcl},  which shows  the charge transfer function for low and high coverages, illustrates the difference more clearly. Here we see that there is no significant difference between Br/Ag(100) and Cl/Ag(100) for $\theta=1/9$, while for $\theta=1/2$ we see a quite significant difference.

\section{Dipole-dipole Interaction}
\noindent In the previous section we have shown that once we have obtained the charge transfer function, we can calculate the dipole moment $p$ from Eq.~(\ref{surface_dip}).
Kohn and Lau~\cite{lau} showed that the non-oscillatory part of the dipole-dipole interaction energy between the adatoms behaves as
\begin{equation}
\phi_{\rm dip-dip}=\frac{2p_Ap_B}{4\pi\epsilon_0R^3}.
\label{dip_dip}
\end{equation}

\noindent The novel aspect of this expression is the factor of 2. A qualitative explanation for this factor is given in Appendix B. For a more detailed and general treatment we refer the reader to Ref.~\cite{lau}. With Eq.~(\ref{dip_dip}), 
we can calculate $\phi_{\rm{dip-dip}}$ from the surface dipole moment results 
from the DFT as described in Eq.~(\ref{surface_dip}) as
\begin{equation}
\phi_{\rm{dip-dip~nnn}}=\frac{2p^2}{4\pi\epsilon_0R^3_{\rm{nnn}}},
\label{dftdip}
\end{equation}
for large $ R $ (in our case larger than the nearest-neighbor distance). Here $p$ is the surface dipole moment calculated from the charge transfer function (Eq.~(\ref{surface_dip})), and $R_{\rm{nnn}}$ is the lateral distance between a pair of next-nearest neighbor adatoms. 

\section{Lattice-gas Model}

\noindent We use an $L \times L$ square array of $N_{\rm site}=L^2$ adsorption sites. Each  site corresponds to a four-fold hollow site on the Ag(100) surface. The energy of this {\rm lattice-gas model} is

\begin{eqnarray}
E=-\sum_{i<j} \phi_{ij} c_i c_j - E_{\rm{b}} \sum_i^{N_{\rm site}} c_i.
\label{lg1}
\end{eqnarray}
Here $i$ and $j$ denote adsorption sites, $\phi_{ij}$ is the lateral interaction energy of the pair ($ij$), and $E_{\rm{b}}$ is the single-particle  binding energy. The sign convention is that $\phi_{ij} < 0$  signifies repulsive interaction and $E_{\rm{b}} >0$ favors adsorption~\cite{mit}. $\Sigma_{i<j}$ is a sum over all pairs of  sites, and $N_{\rm site}$ is the number of four-fold hollow  sites on each side of the slab. 
For simplicity we ignore multiparticle interactions \cite{STAS06,TIWA07}. 

Koper~\cite{kop} has shown that the effects of screening and  finite nearest-neighbor repulsion  are very small. Following his results, we use a lattice-gas model with nearest-neighbor exclusion and unscreened dipole-dipole interactions. The distances used in the lattice-gas model are $R_{ij}=r_{ij}a$ and $R_{\rm nnn}=\sqrt{2}a$, where $R_{ij}$ is the distance between a pair of adsorbates $ij$, and $a$ is the Ag(100) lattice spacing. We can then write
\begin{eqnarray}
\phi_{ij}=\frac{R_{\rm nnn}^3}{R_{ij}^3}\phi_{\rm nnn}
=\frac{(\sqrt{2})^3}{r_{ij}^3}\phi_{\rm nnn}.
\label{li}
\end{eqnarray}

\noindent Thus we have 
\begin{eqnarray}
\frac{1}{N_{\rm site}}\sum_{i<j}\phi_{ij}c_ic_j =\phi_{\rm nnn}\Sigma_{\theta}, 
\end{eqnarray}
where
\begin{eqnarray}
\Sigma_{\theta}=\frac{(\sqrt{2})^3}{N_{\rm site}}\sum_{i<j} \frac{c_ic_j}{r_{ij}^3}.
\label{sigmatheta}
\end{eqnarray}
\noindent The adsorption energy defined in Eq.~(\ref{eads1}) is related to the lattice-gas energy of Eq.~(\ref{lg1}) as
\begin{eqnarray}
E_{\rm ads}=\frac{E}{N_{\rm site}}.
\label{sigma}
\end{eqnarray}
This enables us to break down $E_{\rm ads}$ into its lateral-interaction and single-atom binding parts as follows,
\begin{equation}
E_{\rm ads}=-\phi_{\rm nnn}\Sigma_{\theta}-E_{\rm b} \theta,
\label{eads}
\end{equation}
where $\theta$ is the  coverage  (Eq.~(\ref{cov})) as before.  
The subscript $\theta$ in $\Sigma_{\theta}$ signifies that the lateral 
interaction energy is coverage dependent.

\begin{figure}[t]
\scalebox{0.6}{\includegraphics{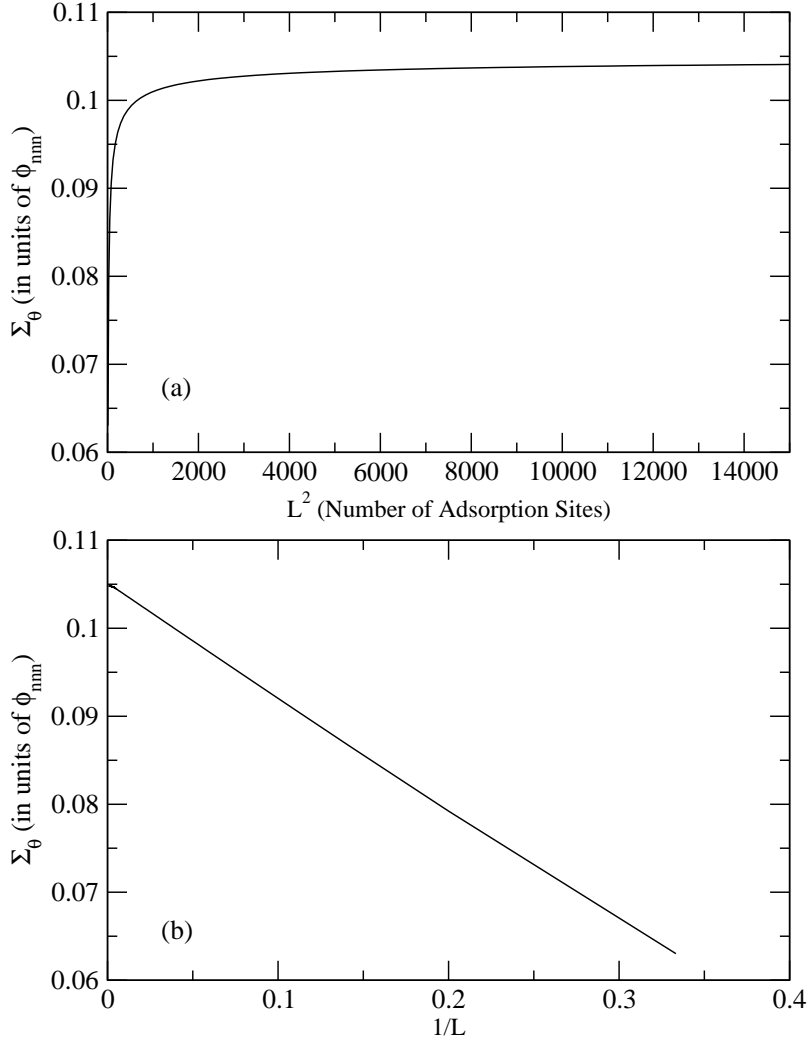}}
\caption{(a)The lateral energy per supercell as a 
function of the number of adsorption sites ($N_{\rm site}=L^2$). As $L^2$ is increased, 
the lateral energy approaches an asymptotic value that can be found by 
plotting the energy per supercell as a function of $1/L$ and extrapolating 
the graph to $1/L = 0$ as shown in (b).}
\label{lat}
\end{figure}

Using the supercell set-up of the DFT,  the lateral part of Eq.~(\ref{lg1}) will be the lateral interaction energy per supercell surface. We can calculate this energy by extending the supercell to infinity in the $x$ and $y$ directions by means of periodic boundary conditions. The central supercell is the original supercell, and the image supercells are the supercell extensions in the $x$ and $y$ directions. The lateral energy per supercell is the sum of the interaction energies of pairs in the central supercell and the lateral energies of pairs of adsorbates in the central supercell and adsorbates in the image supercells. Figure~\ref{lat}(a) shows an example of the lateral energy calculation for $\theta = 1/9$ for finite  $N_{\rm site}$.

The  lateral energy per supercell can be written as
\begin{equation}
\Sigma_{\theta} =\sum^{N_{\rm site}}\frac{C}{r^3},
\label{latsum}
\end{equation}
where $C$ is an arbitrary constant. The above sum can be approximated by the integral
\begin{equation}
\Sigma_{\theta}(L) \approx \int_0^{2\pi} \int_{L_0}^{L}\frac{C}{r^3}rdrd\theta,
\label{latap0}
\end{equation}
which gives us
\begin{equation}
\Sigma_{\theta}(L) \approx \frac{C_1}{L}+C_2 \;.
\label{latap}
\end{equation}
We therefore plot $\Sigma_{\theta}$ versus $ 1/L $.  
It is shown in Fig.~\ref{lat}(b) that the plot is linear in accordance with Eq.~(\ref{latap}).
The correct lateral energy per supercell can then be obtained by fitting Eq.~(\ref{latap}) to the $\Sigma_{\theta}$ versus $ 1/L $ plot and extrapolating  to  $ 1/L = 0 $. The results of this calculations for the different coverages are presented in Table~\ref{tablat}.

\begin{table}[t]
\centering
\caption{\label{tablat} Lateral interaction energy calculation extrapolated to $L \to \infty$. 
$\Sigma_{\theta}$ is the full lateral interaction energy in units of $\phi_{\rm nnn}$, the lateral energy between a pair of next-nearest neighbors.}
\begin{ruledtabular}
\begin{tabular}{cccc}
\hline
$\theta$  & $\Sigma_{\theta}(L \rightarrow \infty)$\\
\hline
1/9 &    0.10512\\
 2/9 &    0.58591\\
  1/4 &    0.79822\\
  1/3 &    1.53990\\
  1/2 &    4.26730\\
\hline
\end{tabular}
\end{ruledtabular}
\end{table}

\section{Lattice-gas Fitting }

\noindent According to our assumption,  $\phi_{ij}$ is quadratic in $p$ and $\sim 1/r^3$. The $\sim 1/r^3$ part has already been calculated in $\Sigma_{\theta}$ as described in Eqs.~(\ref{sigmatheta}) and (\ref{latsum}-\ref{latap}). We also know from the DFT results that the dipole moment $p$ is approximately linear in $\theta$ as shown in 
Fig.~\ref{dipm}. Hence, based on Eq.~(\ref{dftdip}) it is reasonable to assume that we can write $\phi_{\rm{nnn}}$ as~\cite{hamad:1}

\begin{equation}
\phi_{\rm{nnn}}=A(1+B\theta)^2
\label{fit}
\end{equation}

From Eqs.~(\ref{fit})~and~(\ref{eads}), we have  three parameters to be extracted: $A$, $B$, and $E_{\rm b}$. In Fig.~\ref{adse} it is shown that $E_{\rm ads}$ vs $\theta$ is predominantly linear. The linear part is proportional to  
$E_{\rm b}$. The lateral energies contribute to the nonlinear parts which are 
much weaker, and therefore difficult to estimate accurately from a direct 
three-parameter fit. We therefore used the following two-step procedure. 
As can be seen in fig.~\ref{adse}, the graphs extrapolate to 
$E_{\rm ads}(\theta=0)=0$, consistent with the fact that at a very low 
coverage the lateral energy approaches zero. To obtain the dominant 
linear coefficient $E_{\rm b}$, we first fit a quadratic equation 
to $E_{\rm ads}(\theta)$,

\begin{equation}
E_{\rm ads}(\theta) = a_0+a_1\theta+a_2\theta^2.
\label{quad}
\end{equation}

\noindent We extracted the linear part $a_0+a_1\theta$ and used $a_1$ as 
our estimate for the linear coefficient $E_{\rm b}$, 
finding $E_{\rm b}=3.059  \pm 0.058$ eV for Bromine and $E_{\rm b}=3.371 \pm 0.058$ eV for Chlorine. 
We then fixed $E_{\rm b}$ in Eq.~(\ref{eads}) and applied a two-parameter fit to 
extract $A$ and $B$,  which enabled us to calculate $\phi_{\rm{nnn}}$.
(The parameters $a_0$ and $a_2$ are complicated functions of $A$ and $B$ and 
were discarded in favor of the direct two-parameter fit of the latter.)

Using $E_{\rm b}$ from above, we calculate the contribution of the lateral interactions to $E_{\rm ads}$ as
\begin{equation}
C(\theta)=E_{\rm ads}(\theta)+E_{\rm b}\theta.
\label{curvature}
\end{equation}

\begin{figure}[t]
\scalebox{0.6}{\includegraphics{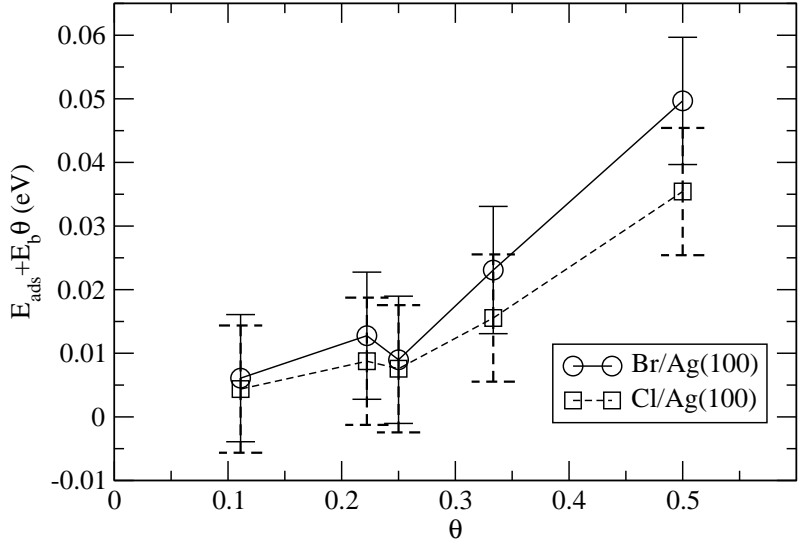}}
\caption{The contributions of the lateral interactions to  $E_{\rm ads}$, shown vs $\theta$.}
\label{curv}
\end{figure}

\begin{figure}[t]
\scalebox{0.6}{\includegraphics{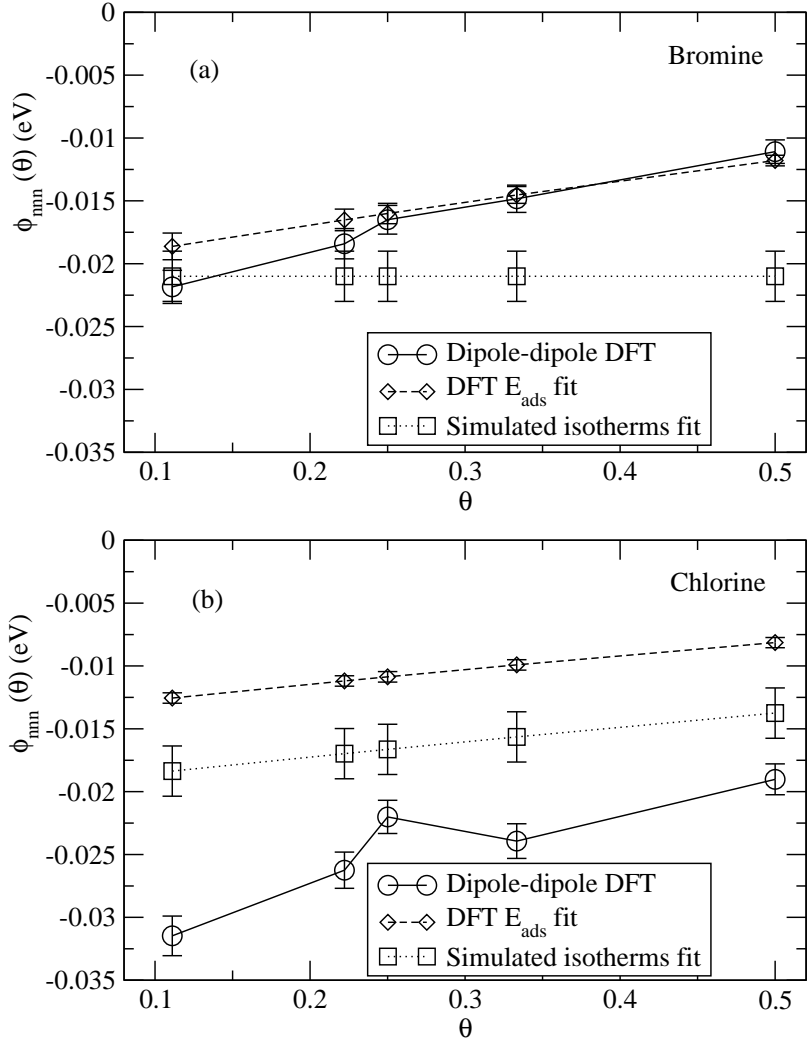}}
\caption{Comparison of the estimates of 
$\phi_{\rm nnn}$ from lattice-gas fit to the adsorption energies (diamonds) 
and from the dipole-dipole interactions (circles). 
Also shown are the results of fitting room-temperature Monte Carlo simulations 
to electrochemical adsorption isotherms from Refs.~\cite{hamad:2,hamad:1} 
(squares). 
(a): Bromine. (b): Chlorine. 
}
\label{fitting}
\end{figure}

\noindent In Fig.~\ref{curv} we plot Eq.~(\ref{curvature}). From this figure it is obvious that 
the lateral energy terms are important.
Figure~\ref{fitting} shows the fitting results for $\phi_{\rm nnn}$. 
It is shown in the figure that for Br the lattice-gas model
obtained by fitting to the adsorption energies  from the DFT
calculation is
consistent with long-range  dipole-dipole lateral interactions using the
dipole moments 
calculated from the DFT charge distribution. This indicates that
long-range dipole-dipole interactions are dominant in this system. 
For Cl the figures show that the long-range dipole-dipole interactions are 
important but not dominant.

We further note that for low coverages our estimates of $\phi_{\rm nnn}$ 
for Br are in excellent agreement  
with those obtained by fitting Monte Carlo simulation results for the lattice-gas model to 
electrochemical adsorption isotherms in Ref.~\cite{hamad:2}.  
However, the DFT results show a stronger coverage dependence than obtained 
from the experimental Monte Carlo fits.  
The experimental fitting results for Cl from Ref.~\cite{hamad:1} lie between 
the two DFT estimates, and all three results show approximately the same 
coverage dependence.

\section{Discussion}
\noindent The lattice-gas model in our study consists of two terms, the lateral interaction term and the single-atom binding-energy term. By fitting the lattice-gas model to adsorption energies obtained from DFT calculations, we have calculated the total lateral energy of the systems. From the charge distribution results from DFT, we have calculated the long-range dipole-dipole interaction contribution to the lateral energy terms that falls off as 
$ \sim 1/r^3$. With this assumption, we calculated dipole-dipole lateral 
interactions by Eq.~(\ref{dip_dip}).

Apart from the difference of magnitude of the dipole moments between Br/Ag(100) and Cl/Ag(100), we find that there are differences in the charge distribution around the adsorbates between Bromine and Chlorine. This is an indication that there are important differences between Br/Ag(100) and Cl/Ag(100).

For Bromine, we showed that the lateral energy calculations from the DFT charge distributions are consistent with the results from fitting the lattice-gas model to the DFT adsorption energies. This shows that in the case of Bromine the lateral energy terms are dominated by long range dipole-dipole interactions. In the case of Chlorine, the lateral energy results from the charge distributions are greater in magnitude than those of Bromine, showing that the long-range dipole-dipole interaction in Cl/Ag(100) is important. However, in the case of Chlorine, we see less consistency between the two methods of calculations. This indicates the presence of significant short-range interactions.

Our calculations were done in vacuum. 
We note, however, the overall consistency of the vacuum DFT calculations 
presented here with previous fits of lattice-gas Monte Carlo simulations to 
electrochemical adsorption isotherms. This suggests that our calculations might be useful to understand these experimental results, in which water is present, as well.

\section*{Acknowledgments}

P.A.R.\ dedicates this paper to his long-time friend and collaborator, 
Andrzej Wieckowski, on the occasion of his 65th birthday.

This work was supported in part by U.S.\ National Science Foundation 
Grant No.\ DMR-0802288 and by The Center for Materials Research and 
Technology (MARTECH) at Florida State University, and by U.S.\ 
Department of Energy Contract No.\ DE-SC0004600 at The University of Tulsa. 
The DFT calculations were performed at Florida State University's 
High-Performance Computing Center. 

\appendix
\setcounter{table}{0}
\section{Convergence Checks}

The number of metal layers in our DFT simulation was determined by  convergence checks.
We calculated $E_{\rm ads}$ for $\theta=1/9$ and $1/2$, for 5, 7, and 9 layers with exactly the same simulation parameter set-up (energy cutoff, $k$-points, the thickness of the vacuum regions, etc.). From Table~\ref{tablat1} we see that increasing the number of metal layers from 5 to 7 changed   $E_{\rm ads}$ for Bromine by less than 1 meV for $\theta=1/9$ and less than 10 meV for $\theta=1/2$. Similar observations are also shown in Table~\ref{tablat2} for Chlorine. Increasing the number of metal layers from 5 to 7, changed $E_{\rm ads}$ for Chlorine by less than 2 meV for $\theta=1/9$ and less than 10 meV for $\theta=1/2$. 

We also calculated the surface dipole moments for  $\theta=1/9$, and $1/2$, for 5, 7, and 9 layers from the above simulations. The convergence check for dipole moments as shown in Table~\ref{tablat3} and \ref{tablat4} shows that increasing the number of layers from 7 to 9 did not change the dipole moment significantly. 

We calculated the percent errors, defined as follows

\begin{equation}
PE_E =\left| \frac{E_{\rm ads}(i)-E_{\rm ads}(j)}{E_{\rm ads}(j)} \right| \times 100\%, 
\label{pe1}
\end{equation}

\begin{equation}
PE_p =\left| \frac{p(i)-p(j)}{p(j)} \right| \times 100\%.
\label{pe2}
\end{equation}

\noindent Here, $PE_E$ is the percent error for adsorption energies $E_{\rm ads}$ and $PE_P$ is 
the percent error for surface dipole moments. In our calculation $j=5$ represents the slab with 5 layers, and $i=7,9$ represent the slabs with 7 and 9 layers, respectively. These percent errors are also shown  in Tables~\ref{tablat1}-\ref{tablat4}

From these two convergence checks ($E_{\rm ads}$ and $p$) we concluded that we need at the very least 5 layers of metal, and  we decided to use 7 layers. Taking the highest value of $E_{\rm ads}(i)-E_{\rm ads}(i-2)$ from 5 to 7 layers, which is 7 meV, we estimate the error bars for 
$E_{\rm ads}$ to be $\Delta E_{\rm ads}= \pm 10$ meV and for $p$ to be $\Delta p= \pm 0.01 e{\rm \AA}$. 
 Error-bar estimates  for $\phi_{\rm nnn}$ based on $\Delta p$ were then calculated by direct error propagation. Error-bar estimates for $E_{\rm ads}$ were obtained as those leading to a $10 \%$ increase in the $\chi^2$ of the two-parameter fit.

\begin{table}
\caption{\label{tablat1} Convergence check for the Bromine adsorption energy 
(in units of eV) with respect to the number of metal layers.}
\begin{ruledtabular}
\begin{tabular}{ccccc}
BROMINE\\
Metal Layers & Coverage & $E_{\rm ads}$    & $E_{\rm ads}(i)-E_{\rm ads}(i-2)$ & $PE_E$\\
\hline
\hline
5      & 1/9      & $-$0.334187984 &   ---  &\\
7      & 1/9      & $-$0.333754808 &     0.0004331 & 0.13\\
9      & 1/9      & $-$0.339772195 &    $-$0.00601 & 1.67\\
\hline
5      & 1/2      & $-$1.475041628 &      --- &\\
7      & 1/2      & $-$1.479772329 &     0.00473 & 0.32\\
9      & 1/2      & $-$1.489228380 &    $-$0.00946 & 0.96\\

\end{tabular}
\end{ruledtabular}

\end{table}

\begin{table}
\caption{\label{tablat2}Convergence check for the Chlorine adsorption energy 
(in units of eV) with respect to the number of metal layers.}

\begin{ruledtabular}
\begin{tabular}{ccccc}
CHLORINE\\
Layers & Coverage & $E_{\rm ads}$    & $E_{\rm ads}(i)-E_{\rm ads}(i-2)$ & $PE_E$\\
\hline
\hline
5     & 1/9     & $-$0.371225625  &      ---  & --- \\
7     & 1/9     & $-$0.370121449  &  0.001104  & 0.29\\
9     & 1/9     & $-$0.376717001  & $-$0.005491  & 1.48\\
\hline
5     & 1/2     & $-$1.642027259  &     ---    & ---\\
7     & 1/2     & $-$1.649943352  & $-$0.007916  & 0.48\\
9     & 1/2     & $-$1.730707884  & $-$0.080765  & 5.40\\
\end{tabular}
\end{ruledtabular}
\end{table}

\begin{table}
\caption{\label{tablat3} Convergence check for the Bromine surface dipole moment 
(in units of ${\rm e\AA}$) with respect to the number of metal layers.}
\begin{ruledtabular}
\begin{tabular}{ccccc}
BROMINE\\
Layers & Coverage &  $p$    & $p(i)-p(i-2)$ & $PE_p$\\
\hline
\hline
5 &  1/9      &   $-$0.241532 &  ---  & ---\\
7 &  1/9      &   $-$0.166162 & 0.075369 & 31.20\\
9 &  1/9      &   $-$0.178799 & $-$0.012637 & 25.97\\
\hline
5 &  1/2      &   $-$0.124252 & --- & ---\\
7 &  1/2      &   $-$0.118376 & 0.005876 & 4.72\\
9 &  1/2      &   $-$0.120788 & $-$0.002412 & 2.78\\
\end{tabular}
\end{ruledtabular}
\end{table}

\begin{table}
\caption{\label{tablat4} Convergence check for the Chlorine surface dipole moment 
(in units of ${\rm e\AA}$) with respect to the number of metal layers}
\begin{ruledtabular}
\begin{tabular}{ccccc}
CHLORINE\\
Layers & Coverage &  $p$    & $p(i)-p(i-2)$ & $PE_p$\\
\hline
\hline
5  & 1/9      &   $-$0.267334 &  --- & ---\\
7  & 1/9      &   $-$0.199436 & 0.067898 & 25.39\\
9  & 1/9      &   $-$0.209752 & $-$0.010316 & 21.54\\
\hline
5  & 1/2      &   $-$0.148922 &  --- \\
7  & 1/2      &   $-$0.155043 & $-$0.006121 & 4.11\\
9  & 1/2      &   $-$0.151803 & 0.003239 & 1.93\\
\end{tabular}
\end{ruledtabular}
\end{table}

\section{The Factor 2 in Eq.~(\ref{dip_dip})}
Following Ref.~\cite{lau}, a qualitative explanation for the factor 2 in Eq.~(\ref{dip_dip}) can be obtained as follows.
Consider an adatom A with induced charge $q_A$, at a distance $z_A$ above the plane surface of a semi-infinite conducting medium, located at $z=0$. The charge-transfer function, integrated over $x$ and $y$, is

\begin{equation}
\Delta\rho_A(z)=-q_A\delta(z)+q_A\delta(z-z_A),
\label{ctf1}
\end{equation}

\noindent where $\delta(z)$ is the Dirac delta function. This yields the dipole moment,

\begin{equation}
p_A=\int_{-\infty}^{\infty}z\Delta\rho(z)dz=-q_A\cdotp0 + q_Az_A=q_Az_A
\label{ctf2}
\end{equation}

\noindent This is the physical dipole created by adatom A. However, the electrostatic potential at a point $z_B$, a lateral distance $R>>z_A$ from $A$, is that of the dipole formed by $q_A$  and its image charge $-q_A$ at $z=-z_A$,

\begin{equation}
U_A(z_B,R)=2z_Aq_A\frac{z_B}{4\pi\epsilon_0R^3},
\label{ctf3}
\end{equation}

\noindent for $z_B\geq0$. This is  equivalent to the potential of a fictitious dipole of magnitude $2z_Aq_A=2p_A$, twice the magnitude of the physical dipole in Eq.~(\ref{ctf2}).

An adatom B with induced charge $q_B$ at $z_B$ corresponds to the  charge transfer function

\begin{equation}
\Delta\rho_B(z)=-q_B\delta(z)+q_B\delta(z-z_B),
\label{ctf4}
\end{equation}
\noindent which gives $p_B=q_Bz_B$.

The potential energy of the pair of adatoms is then

\begin{align}
U_{AB} &= \int_{-\infty}^{+\infty}U_A(z,R)\Delta\rho_B(z)dz \nonumber
\\ &= -q_BU_A(0,R)+q_BU_A(z_B,R) \nonumber
\\ &= 0+\frac{2z_Aq_Az_Bq_B}{4\pi\epsilon_0R^3} \nonumber
\\ &= \frac{2p_Ap_B}{4\pi\epsilon_0R^3},
\label{ctf6}
\end{align}

\noindent which is Eq.~(\ref{dip_dip}). In Ref.~\cite{lau} it is shown that this result holds in more general situations as well, such as jellium and crystalline metals.

\end{document}